\documentclass[a4paper,11pt]{article}
\usepackage{slashed}

\usepackage{braket}

\pdfoutput=1 

\usepackage{jheppub} 

\usepackage[T1]{fontenc} 

\usepackage{xcolor} 

\title{QED-IR as Topological Quantum Theory of Dressed States}

\author[a,1]{J . Gamboa,\note{Corresponding author}}
\author[a]{F. Méndez,}

\affiliation[a]{Departamento de Física, Universidad de Santiago de Chile,
\\Av. Víctor Jara 3493, Santiago, Chile}

\emailAdd{jorge.gamboa@usach.cl}
\emailAdd{fernando.mendez@usach.cl}

\abstract{  We investigate  quantum  electrodynamics  in the  infrared
  regime (QED-IR) using the  adiabatic approximation and the framework
  of the functional Berry phase.  In this approach, the physical state
  space  is  exact,  nonperturbatively  dressed, and  endowed  with  a
  topological structure. Electrons do not exist as bare particles, but
  as topologically  protected electron–photon  clouds, defining  a new
  kind of “infrared quantum”. These  clouds are weakly bound in energy
  --with        a       binding        scale       estimated        at
  \( \Lambda_{\text{IR}}  \sim 0.5~\text{meV}  \)-- and  remain stable
  provided photon energies remain below this threshold. Crucially, the
  theory  becomes  exactly   solvable  in  this  regime   due  to  the
  quantization  of  the  functional  Berry  flux,  which  governs  the
  infrared dynamics of the dressed states.

  {When  hard (high-energy)  processes are  involved, the  topological
    protection  of  the  dressed  states is  lifted,  and  the  theory
    smoothly recovers  conventional perturbative QED. In  contrast, in
    the  deep infrared,  the electromagnetic  interaction never  fully
    vanishes, leading to observable effects.  We argue that the energy
    required to dissolve the infrared electron--photon cloud in QED is
    of order  $\text{meV}$, comparable  to the  thermal energy  of the
    cosmic            microwave           background            (CMB),
    $kT_{\text{CMB}} \approx 2.3  \times 10^{-4}\,\text{eV}$. However,
    the observed  temperature anisotropies correspond  to fluctuations
    near  $10^{-9}\,\text{eV}$, far  too small  to destroy  the cloud,
    though  potentially capable  of perturbing  its topological  phase
    structure.  This  suggests  that   CMB  deviations  could  reflect
    residual   topological  imprints   of   the  functional   infrared
    dynamics. Finally, we propose that analogous cloud-like structures
    may  manifest  in other  quantum  systems  governed by  low-energy
    photon dynamics, such as atomic and molecular environments.}  }

\begin{document} 
\maketitle
\flushbottom


\section{Introduction}

A longstanding problem in quantum  field theory, first raised by Bloch
and  Nordsieck in  1937~\cite{BN}, concerns  the infrared  divergences
that  arise  in theories  involving  massless  gauge bosons,  such  as
quantum electrodynamics (QED). This  issue appears, for instance, when
analyzing the  elastic scattering of  an electron off a  heavy source:
the  electron necessarily  emits  photons of  arbitrarily low  energy,
which remain practically undetectable.

Bloch and  Nordsieck~\cite{BN} showed  that calculating  the amplitude
for  a  process  with  no  soft-photon  emission  yields  a  divergent
result.  However, by  summing  over  all physically  indistinguishable
final   states—including  those   with   soft  photons—the   resulting
probabilities become finite~\cite{YFS}.

In the 1960s, Kinoshita~\cite{KI} and later Lee and Nauenberg~\cite{LN} refined this idea by demonstrating that the cancellation of infrared divergences also requires summing over degenerate \emph{initial} states. This led to the formulation of the KLN theorem, which ensures the finiteness of physical observables in the presence of massive charged particles.

Around the same time, Weinberg~\cite{weinberg} generalized the discussion to include infrared gravitons, reinforcing the idea that infrared divergences can, in principle, be handled systematically. Nonetheless, a certain dissatisfaction remains, as the deeper physical implications of these divergences are still not fully understood.

A major conceptual shift was proposed in the early 1970s by Kulish and Faddeev~\cite{chung,KF,carney1,carney2,ref1,ref2,reclamo1,reclamo2}, who suggested modifying the construction of asymptotic states. In their approach, the \textit{in} and \textit{out} states must be dressed with coherent clouds of infrared photons that encode the entire interaction history of the charges. In effect, this amounts to abandoning the idea that plane waves correctly describe the asymptotic states of QED in the infrared—true physical states no longer belong to Fock space.

This idea introduced a compelling conceptual framework: the low-energy regime of QED is necessarily governed by a non-perturbative vacuum with nontrivial functional or topological structure. The Kulish–Faddeev construction marked a major technical advance, showing that physical states must be composite electron–photon configurations, thereby challenging the standard perturbative framework and anticipating the need for a nontrivial infrared sector.

However, from a theoretical standpoint, the Kulish–Faddeev approach remains somewhat unsatisfying, as the dressing factor is introduced in an \textit{ad hoc} manner, with no underlying principle to determine its form. This motivates the search for a more systematic framework in which the asymptotic dressing arises naturally.

Such a framework would be valuable not only for going beyond Fock space, but also for providing insight into ultra-low-energy physics and the nonperturbative structure of gauge theories~\cite{FS,MorchioStrocchi,StrocchiBook,DerezinskiGerard,
Buchholz1,Buchholz2,Mund,Herbst,Herdegen}.

Inspired by Berry's work~\cite{Berry} and its implications for effective theories, we have proposed~\cite{yo,yo2} that the QED-IR can be analyzed using a variant of the adiabatic approximation. In this framework, an emergent Berry connection acts as a mediator between ultraviolet and infrared regimes, closely paralleling its role in the Born–Oppenheimer approximation of quantum mechanics~\cite{shapere}.

In this context, Berry's insights into adiabatic evolution in quantum mechanics serve as a natural guide: the wavefunction acquires a topological, non-perturbative phase, which may lead to observable physical consequences. This type of structure—more rigid and universal than the arbitrary Kulish–Faddeev dressing—suggests that the dressing of states is governed by a functional Berry phase determined by the topology of configuration space~\cite{yo2}.

From this perspective, QED-IR can be reformulated using the adiabatic approximation, with the accumulated Berry phase providing a non-perturbative dressing of the physical states. This formulation shifts the emphasis from local dynamics to topology: the formal structure of QED remains unchanged, but the boundary conditions in functional space give rise to physically meaningful consequences.

Motivated by these foundational ideas, we revisit the infrared structure of QED from a new perspective inspired by Berry's geometric phase~\cite{Berry}. In this work, we present an alternative formulation that retains the essential elements of the Berry approach while incorporating novel geometric and topological structures.

The key ingredients of our construction are:
\begin{itemize}
    \item[(a)] A variant of the adiabatic approximation applied to the Dirac operator, enabling the construction of an effective action involving both the original gauge fields and the emergent Berry connection.
    \item[(b)] The introduction of a convenient gauge choice—referred to as the \emph{adiabatic gauge}—which fixes the \( U_{\text{Berry}}(1) \) redundancy associated with the adiabatic structure.
\end{itemize}

With these two elements, the construction of dressed states in infrared QED becomes not only natural but also remarkably transparent. These dressed states provide explicit realizations of a non-Fock physical space, labeled by an integer \( n \), which emerges from the functional quantization of the Berry holonomy.

The paper is organized as follows. In Section~II, we present the general framework. Section~III details the implementation of the adiabatic approximation in QED. In Section~IV, we analyze the dressed states and the topological quantization of the functional Berry flux. Section~V discusses the concept of topological binding in electron–photon clouds. In Section~VI, we explore broader implications of functional dressing, including potential connections to the cosmic microwave background (CMB) and the notion of “free” electrons. Finally, Section~VII contains our conclusions.

\section{Developing the Idea}

To develop the ideas outlined above, let us consider the generating functional of QED:
\begin{equation}
\mathcal{Z} = \int [\mathcal{D}A]\, \mathcal{D}\bar{\psi}\, \mathcal{D}\psi \, e^{-S[\bar{\psi}, \psi, A]}, \label{func1}
\end{equation}
where \( [\mathcal{D}A] \) denotes the Faddeev–Popov gauge-fixed measure, and
\begin{equation}
S_0 = \int d^4x \left( \frac{1}{4} F_{\mu\nu} F^{\mu\nu} + \bar{\psi} \slashed{D}[A] \psi \right), \label{sch2}
\end{equation}
with \( D_\mu [A] = \partial_\mu + i e A_\mu \).

\medskip

We will consider massless fermions for the following physical reason: since we are interested in processes dominated by soft photons with momentum \( k \) interacting with electrons of momentum \( p \), the relevant regime satisfies \( |k| \sim |p| \). This condition is naturally realized when the electron mass is negligible compared to its momentum, thus justifying the use of massless fermions.

At this point, it is also worth drawing a parallel with the asymptotic symmetry program developed by Strominger \textit{et al.}~\cite{stro,stro1,stro2,stro3}, where massless fermions are likewise assumed. Although their analysis is based on BMS symmetries and celestial amplitudes, the approximation of massless fermions in the infrared is consistent with the physical setting of soft-photon emission.

\medskip

Let us now perform the local chiral transformation
\begin{equation}
\psi(x) \to e^{i \alpha(x) \gamma_5} \psi(x), \qquad \bar{\psi}(x) \to \bar{\psi}(x)\, e^{i \alpha(x) \gamma_5}, \label{chiral}
\end{equation}
in the functional integral \eqref{func1}. Under this transformation, the action becomes:
\begin{equation}
S_1 = \int d^4x \left( \frac{1}{4} F_{\mu\nu} F^{\mu\nu} + \bar{\psi} \left( \slashed{D}[A] - i \gamma_5 \slashed{\partial} \alpha(x) \right) \psi \right), \label{sch23}
\end{equation}
while the fermionic measure transforms, as shown by Fujikawa~\cite{fujikawa1,fujikawa2}, according to:
\begin{equation}
\mathcal{D}\bar{\psi} \, \mathcal{D}\psi \ \to \ \mathcal{D}\bar{\psi} \, \mathcal{D}\psi \, \exp\left( -\frac{e^2}{16\pi^2} \int d^4x \, \alpha(x)\, F_{\mu\nu} \tilde{F}^{\mu\nu} \right). \label{sch24}
\end{equation}

\medskip

Integrating by parts in \( \alpha(x) \), the condition \( \delta \mathcal{Z} / \delta \alpha(x) = 0 \) yields the chiral anomaly. Collecting all contributions, we obtain the effective action~\cite{yo,yo2}:
\begin{equation}
S_1 = \int d^4x \left( \frac{1}{4} F_{\mu\nu} F^{\mu\nu} + \bar{\psi} \left( \slashed{D}[A] - i \gamma_5 \slashed{\partial} \alpha(x) \right) \psi -\frac{e^2}{16\pi^2}\alpha(x)\, F_{\mu\nu} \tilde{F}^{\mu\nu} \right). \label{sch25}
\end{equation}

This expression, in the spirit of Fujikawa’s method, generates the chiral anomaly. Although this is not the conventional form of the QED action—since anomalies are typically neglected in standard electromagnetic contexts—it proves particularly suitable for exploring the infrared regime, where phases associated with local chiral transformations become physically meaningful.

\medskip

Since our aim is to investigate QED in the infrared using the adiabatic approximation, it is natural to add the analogue of the \( \theta \)-term,
\[
S_\theta = \int d^4x\,\, \theta\, \frac{e^2}{16\pi^2} F_{\mu\nu} \tilde{F}^{\mu\nu},
\]
but promoting \( \theta \) to a spacetime-dependent field \( \theta(x) \), rather than treating it as a constant.

Thus, we consider the modified action:
\begin{equation}
S_\theta = \int d^4x\, \frac{e^2}{16\pi^2}\,\theta(x) F_{\mu\nu} \tilde{F}^{\mu\nu}, \label{stheta}
\end{equation}
and add it to the anomalous action \eqref{sch25}.

Now, if we identify
\[
\theta(x) = \alpha(x),
\]
the topological terms proportional to \( F_{\mu\nu} \tilde{F}^{\mu\nu} \) cancel out, and the resulting action simplifies to:
\begin{equation}
S = \int d^4x \left( \frac{1}{4} F_{\mu \nu} F^{\mu \nu} 
+ \bar{\psi} \left( \slashed{D} -i \gamma_5 \slashed{\partial}\alpha \right) \psi \right). \label{effe1}
\end{equation}

\medskip

At this point, one may wonder whether the field \( \theta(x) \) could be interpreted dynamically, perhaps in analogy with the axion. In our case, however, the identification \( \theta = \alpha \) is introduced specifically to simplify the implementation of the adiabatic approximation, following Berry's geometric framework~\cite{Berry}.

\section{Adiabatic Approximation}

The adiabatic approximation employed in this work is inspired by its standard use in quantum mechanics, where it is typically applied to systems with slowly varying parameters—such as time-dependent Hamiltonians with well-separated energy levels. However, our motivation and implementation differ in significant and essential ways.

First, we do not assume the existence of a preferred time parameter, nor of any slowly varying external field. Instead, our starting point is a modified Dirac operator, arising from a local chiral transformation involving a functional parameter \( \alpha(x) \), whose variation is not assumed to be small or adiabatic in the traditional sense.

Second, rather than focusing on time evolution, our adiabatic ansatz aims to uncover the geometric and topological content encoded in the infrared structure of QED. Specifically, we are interested in how soft photon modes become entangled with the fermionic sector, and how this entanglement is reflected in the spectral properties of the Dirac operator.

Our framework may be viewed as a kind of \emph{functional adiabatic approximation}, where the role of time is replaced by variation along trajectories in configuration space. The key insight is that, even in the absence of physical time evolution, one can still define a geometric phase—analogous to the Berry phase—that accumulates as the system explores the space of soft gauge configurations.

By applying a Berry-inspired ansatz to the eigenvalue problem of the modified Dirac operator, we isolate this geometric phase and identify a corresponding Berry connection. This leads naturally to the emergence of dressed fermionic states and to an effective infrared action in which the Berry connection appears explicitly.

In this sense, our adiabatic approximation is not based on dynamical slowness, but on a structural separation between soft and hard modes, and on the geometric response of fermions to smooth deformations in gauge space. This conceptual shift is crucial for understanding the infrared behavior of QED beyond perturbation theory.

\medskip

We implement this idea as follows. The fermionic fields in~\eqref{effe1} can be formally integrated out, yielding the determinant
\begin{equation}
\det \left( \slashed{D}[A] - i \gamma_5 \slashed{\partial}\alpha \right) = \prod_n \lambda_n, \label{fb1}
\end{equation}
where \( \lambda_n \) are the eigenvalues of the operator
\begin{equation}
\left( \slashed{D}[A] - i \gamma_5 \slashed{\partial}\alpha \right) \varphi_n = \lambda_n \varphi_n. \label{fb2}
\end{equation}

To study this spectrum, we apply a Berry-inspired ansatz to the eigenfunctions:
\begin{equation}
\varphi_n(x) = e^{i \gamma_n(x)}\, \bar{\phi}_n(x), \label{fb3}
\end{equation}
where \( \gamma_n(x) \) is a phase to be determined. This ansatz encodes the structure of the eigenstates in terms of a geometric phase accumulated along a functional path, consistent with the emergence of a Berry connection.

Substituting~\eqref{fb3} into~\eqref{fb2}, we obtain the approximate Dirac equation:
\begin{equation}
\left( \slashed{D}[A] + \slashed{\mathcal{A}}^{nn} - i \gamma_5 \slashed{\partial}\alpha \right) 
\bar{\phi}_n = \lambda_n \bar{\phi}_n, \label{fb4}
\end{equation}
where
\begin{equation}
\mathcal{A}^{nn}_\mu = i \langle \bar{\phi}_n | \partial_\mu | \bar{\phi}_n \rangle \equiv \mathcal{A}_\mu \label{fb5}
\end{equation}
is the Berry connection associated with the eigenmode \( \bar{\phi}_n \). In the spirit of the adiabatic approximation, we neglect the off-diagonal components of the connection (see for example \cite{shapere}).

It is important to emphasize that the operator remains Dirac-like, although the spinor basis has changed. Thus, we can approximate the original determinant~\eqref{fb1} as:
\begin{equation}
\det \left( \slashed{D}[A] - i \gamma_5 \slashed{\partial}\alpha \right)
\approx \det \left( \slashed{D}[A] + \slashed{\mathcal{A}} - i \gamma_5 \slashed{\partial}\alpha \right), \label{fb6}
\end{equation}
where the Berry connection appears explicitly.

We may now reinterpret this determinant as that of a Dirac operator acting on a new set of spinors \( \chi \), related to the original fermions by the geometric transformation implied in~\eqref{fb3}. These spinors satisfy
\[
\chi := e^{-i \gamma_n(x)}\, \varphi_n(x),
\]
and are defined such that the Jacobian of the transformation is trivial (\( \det = 1 \)). The resulting effective action becomes
\begin{equation}
S_{\text{eff}} = \int d^4x \left( \frac{1}{4} F_{\mu \nu} F^{\mu \nu} 
+ \bar{\chi} \left( \slashed{D}[A] + \slashed{\mathcal{A}} - i \gamma_5 \slashed{\partial}\alpha \right) \chi \right). \label{eff12}
\end{equation}

This is the infrared effective action we set out to derive. It retains the usual gauge symmetry of QED, but also exhibits an emergent \( U(1) \) structure associated with the Berry connection. To make this symmetry manifest, we impose the following gauge-fixing condition:
\begin{equation}
\mathcal{A}_\mu = i \gamma_5 \partial_\mu \alpha(x), \label{gau1}
\end{equation}
which we refer to as the \emph{adiabatic gauge}.

In this gauge, the action~\eqref{eff12} reduces to the conventional form:
\begin{equation}
S = \int d^4x \left( \frac{1}{4} F_{\mu\nu} F^{\mu\nu} + \bar{\chi}~ \slashed{D}[A] \chi \right), \label{accion2}
\end{equation}
which appears deceptively simple, since it conceals the topological information encoded in the asymptotic structure of the physical states.

Nevertheless, from the adiabatic gauge condition, we recover the accumulated Berry phase:
\begin{equation}
\Delta \alpha = - i \int dx^\mu\, \gamma_5\, {\cal A}_\mu, \label{acumu}
\end{equation}
which leads to a global, non-perturbative dressing of the asymptotic states through the accumulation of a topological phase:
\begin{equation}
| \text{in} \rangle \rightarrow e^{\Delta \alpha_{\text{in}}} | \text{in} \rangle, \qquad
| \text{out} \rangle \rightarrow e^{\Delta \alpha_{\text{out}}} | \text{out} \rangle. \label{boundary}
\end{equation}

\section{The Nature and Meaning of Infrared Dressing}

The existence of electron–photon clouds is a fundamental consequence of the infrared structure of quantum electrodynamics (QED), where—as previously discussed—the traditional notion of asymptotic states breaks down in a nonperturbative manner. In this regime, infrared photons neither carry sufficient energy to be individually detected, nor do they alter the global quantum numbers of the system, such as total angular momentum.

As a result, the physical state—composed of the electron and its coherent cloud of soft photons—preserves the electron’s intrinsic spin-$\frac{1}{2}$. This raises a natural question: how do infrared photons manifest physically if they cannot be individually observed? The answer lies in their cumulative, nonlocal effects, encoded in topological phases and entangled structures. The effective algebra describing these clouds reflects their composite nature and is organized around the fermionic degrees of freedom that remain observable.

From a strict infrared perspective, there are no truly free photons. Even in the paradigmatic case of the CMB—where the interaction rate with matter is extremely low—photons remain entangled with a dressed vacuum and cannot be fully factorized into independent subsystems. In this context, the notion of a \emph{cloud} acquires a dual interpretation: the \emph{body} is the charged particle (e.g., an electron), while the \emph{soul} is the coherent configuration of soft photon modes that necessarily accompanies it. This inseparable structure not only ensures the nonperturbative cancellation of infrared divergences, but also reveals the topological and entangled nature of the gauge-theoretic vacuum.

With these ideas in mind, we revisit the definition of the dressed states in~\eqref{boundary} and ask how they are explicitly constructed. To this end, we note that the accumulated Berry phase in~\eqref{acumu} must be interpreted as an operator acting on the Hilbert space of physical states.

Adopting the chiral representation for \( \gamma_5 \), the Berry phase becomes
\begin{equation}
\mathrm{Tr}\left[e^{\Delta \alpha}\right] = 2\, \mathrm{Re}\left[\mathrm{Tr}\left(e^{-i \oint dx^\mu\, \mathcal{A}_\mu}\right)\right],
\end{equation}
where the Berry connection \( \mathcal{A}_\mu \) is interpreted as an emergent gauge field in the space of adiabatic states, generated by the interaction between the electron and the infrared photonic background. The appearance of \( \gamma_5 \mathcal{A}_\mu \) in \( \Delta \alpha \) encodes a possible chiral asymmetry in the adiabatic evolution, where topological effects—such as Chern numbers or holonomy—may become relevant.

Consequently, the trace
\begin{equation}
\mathrm{Tr}\left[e^{\Delta \alpha}\right] = 4\, e^{\Delta \alpha} = 2\, \mathrm{Re}\left[\mathrm{Tr}\left(e^{-i \oint dx^\mu\, \mathcal{A}_\mu}\right)\right], \label{holo1}
\end{equation}
can be interpreted as an observable sensitive to the holonomy induced by the adiabatic evolution of a composite electron–photon state. This observable may correspond to physical effects such as anomalous phases, effective couplings, or partial confinement mechanisms in the infrared regime.

The explicit evaluation of~\eqref{holo1} yields
\begin{equation}
e^{\Delta \alpha} = \cos\left( \oint_C dx^\mu\, \mathcal{A}_\mu \right). \label{holo2}
\end{equation}
However, since the dressed cloud carries spin-$\frac{1}{2}$, the exponential factor must satisfy \( e^{\Delta \alpha} = -1 \), which implies the quantization condition
\begin{equation}
\oint_C dx^\mu\, \mathcal{A}_\mu = (2n+1)\pi. \label{holo3}
\end{equation}

This condition is reminiscent of the mod-$2$ index theorem~\cite{Atiyah1,Witten1}, which counts the parity of fermionic zero modes of the Dirac operator. In this context, the nontrivial holonomy \( (-1) \) reflects a topological obstruction to globally trivializing the fermionic bundle and corresponds to a \( \mathbb{Z}_2 \)-valued topological invariant associated with spin-$\frac{1}{2}$ fields. The Berry phase, therefore, captures a discrete topological sector of the infrared theory, protected by the statistics of the dressed fermions.

This result is remarkable: in the infrared regime of QED, the space of physical states consists of a discrete family of functionally dressed states \( \{ |\Psi_n\rangle \} \), each labeled by a quantized functional phase \( \Delta \alpha_n = (2n+1)\pi \). The quantization of this functional flux implies that all physical states—including the one corresponding to \( n = 0 \)—are topologically protected. As a result, the perturbative Fock structure is replaced by a fully topological functional basis, where the stability of states arises from their homotopy class rather than energy considerations.

These states may be viewed as the functional analogues of topological solitons: localized configurations in field space, labeled by a discrete invariant, and stable under continuous deformations.

\section{Topological Binding and the Infrared Well in QED}

The functionally dressed state of the electron in QED-IR can be understood as being trapped in an energy well generated by its surrounding cloud of soft photons.

The physical interpretation of dressing clouds in the infrared regime of QED reveals a subtle yet profound phenomenon. Although the formalism is rooted in gauge theory, the underlying intuition resonates with ideas from condensed matter and statistical physics \cite{Hasan,Qi,Moore,Kane,Berne}. In particular, the energetic structure of infrared dressing displays parallels with marginally bound states, albeit with a topological origin.

In QED, the interaction with soft gauge modes precludes the existence of bare Fock states, replacing them with inseparable, coherent electron–photon configurations. This structure may be interpreted as a form of \emph{functional confinement} in the infrared: the physical state no longer corresponds to an isolated elementary particle, but rather to a dynamical, composite configuration shaped by long-range gauge interactions.

However, the nature of this confinement is fundamentally weak. The infrared cloud is fragile—it can reorganize or dissipate under small energetic perturbations. Although dressing is essential for the cancellation of infrared divergences and for preserving gauge invariance, the associated binding energy is nearly zero.

From a geometric perspective, the functional Berry phase accumulated in QED-IR defines a smooth structure without deep topological wells. As such, the dressing corresponds not to a stable topological soliton, but rather to a \emph{marginally bound} quantum state: stable under adiabatic evolution but sensitive to energetic disturbances.

Whether one regards the infrared regime as a technical limit or a genuine physical sector, a natural question arises: \emph{What is the characteristic energy required to dissociate a dressed state and restore the perturbative behavior of QED?}

To address this, we recall that in QED-IR, physical states consist of electrons accompanied by coherent clouds of soft photons. This cloud defines a nontrivial structure that ensures gauge invariance, cancels infrared divergences, and precludes the existence of truly free electrons. Heuristically, one may interpret this configuration as an effective potential well, whose depth corresponds to the minimum energy needed to unbind the cloud and return to standard Fock states.

{
The energy stored in the soft-photon dressing can be estimated using the Kulish--Faddeev profile,
\begin{equation}
f_\lambda(\mathbf k)=-\,e\,\frac{p\!\cdot\!\varepsilon_\lambda(\mathbf k)}{p\!\cdot\! k}\,
\Theta(\omega_k-\epsilon)\,\Theta(\kappa-\omega_k),
\end{equation}
which leads to
\begin{equation}
\Delta E_{\text{cloud}}
=\sum_{\lambda}\int \frac{d^3k}{(2\pi)^3}\,\omega_k\,|f_\lambda(\mathbf k)|^2.
\end{equation}
After the polarization sum and angular average, one finds a velocity factor $\propto\gamma^2=(1-v^2)^{-1}$, while the frequency integral reduces to a finite soft window. The result is
\begin{equation}
\Delta E_{\text{cloud}}
\simeq \frac{2\alpha}{\pi}\,\gamma^2\,(\kappa-\epsilon),
\end{equation}
which is dimensionally consistent and shows the expected linear dependence on the infrared window.

For representative values,
$\epsilon=10^{-3}\,\mathrm{eV}$ (microwave resolution, $\nu\sim240$~GHz) and
$\kappa=0.1\,\mathrm{eV}$, one finds
$\Delta E_{\text{cloud}}\sim 0.5~\mathrm{meV}$ for a nonrelativistic electron.
This demonstrates that the infrared binding in QED is extremely weak:
even modest perturbations suffice to dismantle the dressed state.
}

Thus the infrared “binding” in QED is exceedingly small: comparatively modest energy injections readily disrupt the dressed state. This energetic shallowness explains why topologically nontrivial infrared dressings do not entail confinement—transitions to standard Fock states are easily induced by mild perturbations.

Thus the infrared “binding” in QED is exceedingly small: comparatively modest energy injections readily disrupt the dressed state. This energetic shallowness explains why topologically nontrivial infrared dressings do not entail confinement—transitions to standard Fock states are easily induced by mild perturbations.

\section{Electron--Photon Clouds}

Electron--photon clouds represent the fundamental quantum of infrared QED, and their presence should manifest in concrete physical phenomena. Below, we present two illustrative examples that highlight the physical implications of this perspective.

\subsection{Blackbody Radiation}

{
A natural question arises when comparing the functional infrared description of QED with observational phenomena such as the CMB: if the true infrared degrees of freedom are functionally dressed electron--photon clouds with fermionic structure, how can the CMB be so accurately described as a gas of free photons?

The resolution lies in the energy scales involved. The characteristic binding energy of a dressed electron--photon cloud in QED--IR can be estimated as
\begin{equation}
\Delta E_{\text{cloud}} \;\simeq\; \frac{2\alpha}{\pi}\,\gamma^2\,(\kappa-\epsilon),
\end{equation}
with $\epsilon$ an infrared cutoff (detector resolution) and $\kappa$ a soft factorization scale delimiting the dressing. 
For representative values, $\epsilon\simeq 10^{-3}\,\mathrm{eV}$ and $\kappa\simeq 0.1\,\mathrm{eV}$, one obtains
\[
\Delta E_{\text{cloud}} \sim 5\times 10^{-4}~\text{eV} \quad (0.5~\text{meV}).
\]

By comparison, the temperature of the CMB is
\[
T_{\text{CMB}} \approx 2.725~\text{K} \quad \Rightarrow \quad kT_{\text{CMB}} \approx 2.35 \times 10^{-4}~\text{eV}.
\]
Thus the thermal scale of the CMB is of the same order as the infrared binding energy of the cloud, implying that dressed states are only marginally stable against perturbations at this temperature.

Consequently, the relevant degrees of freedom at the CMB scale are not bare photons, but functionally dressed collective excitations that remain bound at leading order. These dressed states reproduce the infrared dynamics and, despite their fermionic structure, their statistical distribution can effectively mimic that of a bosonic gas.

Nevertheless, the spectral distribution of the CMB follows Planck’s law to high precision \cite{Fixsen:1996,WMAP:2013,Planck:2018}. This is not contradictory: the CMB measures statistical properties of collective excitations, rather than the microscopic nature of the quanta. While functional dressing modifies the vacuum structure and the density of states, the excitations can still follow an effective Bose--Einstein distribution at finite temperature.

However, functional electron--photon clouds may induce small deviations from the ideal blackbody spectrum, especially in the deep infrared regime where dressing effects dominate. To model this, we dress the radiation modes as
\begin{equation}
|n_\nu\rangle_{\text{phys}} = e^{i \Delta \alpha(\nu)} |n_\nu\rangle,
\end{equation}
where $\Delta \alpha(\nu)$ is a functional Berry phase along soft gauge directions. A typical parametrization is
\begin{equation}
\Delta \alpha(\nu) = \frac{2\pi \nu}{\Lambda_{\text{IR}}} + \phi,
\end{equation}
with $\Lambda_{\text{IR}}$ an infrared scale (expected to satisfy $\Lambda_{\text{IR}} \lesssim 10^{-3}\text{--}10^{-2}\,\text{eV}$) and $\phi$ a constant global phase.

The resulting spectral energy density takes the form
\begin{equation}
\rho(\nu) \;\simeq\; \rho_{\text{Planck}}(\nu)\,
\Big[\,1 + \epsilon \cos\!\Big(\tfrac{2\pi \nu}{\Lambda_{\text{IR}}} + \phi\Big)\Big],
\end{equation}
where
\[
\rho_{\text{Planck}}(\nu) = \frac{8\pi h \nu^3}{c^3}\,\frac{1}{e^{h\nu/kT}-1}.
\]

The relative deviation from the ideal spectrum is then bounded by
\begin{equation}
\left|\frac{\rho - \rho_{\text{Planck}}}{\rho_{\text{Planck}}}\right| \leq \epsilon.
\end{equation}

Since a small temperature shift $T\to T+\delta T$ produces
\[
\left|\frac{\rho - \rho_{\text{Planck}}}{\rho_{\text{Planck}}}\right| \sim \frac{\delta T}{T},
\]
the functional deviation must satisfy
\begin{equation}
\epsilon \lesssim \frac{\delta T}{T} \sim 10^{-5},
\end{equation}
in agreement with observational constraints on CMB anisotropies \cite{Planck:2018}. This coincidence suggests that possible deviations from blackbody behavior may have a topological origin linked to infrared dressing, rather than to ordinary thermodynamic fluctuations. 

While the remarkable agreement of the global CMB spectrum with Planck’s law constrains any deviations to the level of $10^{-5}$, this bound does not exclude the existence of subleading topological imprints in specific frequency windows. In particular, in the Rayleigh--Jeans tail ($h\nu \ll kT$), the spectrum is especially sensitive to small distortions, and residual analyses in the $20$--$50$~GHz range may already be probing the relevant regime \cite{QUIJOTE:2020,QUIJOTE:2023}. Current experiments such as QUIJOTE, together with complementary low-frequency measurements from WMAP and Planck \cite{WMAP:2013,Planck:2018}, therefore provide a natural arena to test for functional dressing effects. In this way, the two perspectives are consistent: global CMB observations impose stringent bounds, while targeted Rayleigh--Jeans data may reveal the topological origin of small departures from the ideal blackbody spectrum \cite{GamboaRJ}.

}

\subsection{The Myth of the Free Electron in the Infrared Regime}

The notion of functional clouds forces us to reconsider foundational assumptions that are often taken for granted. One prominent example is the concept of a free electron gas. Such a gas does not exist as an ontological entity, because in the infrared regime of QED, electrons do not exist as physical, bare states. The free electron is not an observable; only functionally dressed clouds exist, whose topological structure forbids the existence of isolated electrons.

Unlike the standard perturbative approach—where soft photons are treated as negligible corrections—the functional formulation of infrared QED reveals that the soft sector defines the physical state itself. The dressed configuration, composed of an electron and its soft photon cloud, remains dynamically relevant even when the photon energy is non-relativistic. That is, dressing does not disappear at low energies; instead, it persists as long as the photons remain below a characteristic threshold.

This threshold, denoted by \( \Lambda_{\text{IR}} \), can be interpreted as the depth of the functional binding well. A typical estimate gives \( \Lambda_{\text{IR}} \sim 0.5~\text{meV} \), representing the minimal energy required to dissolve the cloud and transition to a regime where the electron may be treated as free. As a result, infrared QED predicts that even slow-moving electrons—such as those found in condensed matter systems or atomic bound states—cannot be treated as bare particles. They are always accompanied by a nontrivial functional phase, unless their interactions with photons involve energies above \( \Lambda_{\text{IR}} \).

\section{Final Comments and Outlook}

A central conclusion of this work is that the dressed states of quantum electrodynamics in the infrared regime constitute exact and topologically protected solutions, in which the electron--photon cloud emerges as the fundamental excitation---the effective ``quantum'' of the infrared sector. This perspective transcends the reach of conventional perturbative methods and exposes a structure that is at once conceptually simple and physically profound, carrying significant implications for our understanding of infrared dynamics and the ontology of particles in gauge theories.  

One of the key insights developed here is that dressed clouds represent topologically stable configurations in the infrared domain of QED. Interpreted as genuine ontological entities, these configurations dictate the low-energy dynamics and endow the space of physical states with a nontrivial topological structure. Their stability also suggests a possible role in systems traditionally regarded as free, including the photons of the cosmic microwave background (CMB).  

Within this framework, the observed temperature fluctuations in the CMB can be viewed as consistent with the minimal energy scale required to stabilize the cloud. In this sense, such fluctuations may not be reducible to mere thermodynamic noise, but could represent subtle signatures of infrared QED effects encoded in the topological properties of dressed states.  

The relevance of dressed clouds, however, is not confined to the infrared regime of QED. More broadly, this concept offers a novel perspective on fundamental problems in quantum mechanics wherever electromagnetic radiation plays a central role---ranging from atomic and molecular physics to many-body systems. While conventional approaches often rely on perturbative expansions and bare particle states, our findings indicate that whenever soft-photon exchanges accumulate coherently, a dressed-state formulation provides a more accurate and conceptually robust description.  

This shift in viewpoint---from bare particles to dressed clouds, from perturbative series to topological structures---opens a promising path for reinterpreting the foundations of quantum field theory. It highlights the possibility that the true building blocks of gauge theories are not bare quanta, but topologically protected dressed states that redefine the very nature of physical excitations.  

{The idea of analyzing IR-QED from the perspective of the Berry phase provides an economical explanation of a problem that has not previously been discussed in this manner, and sheds light on a non-perturbative aspect that is also not well explored. The underlying mathematical structure \cite{Simon}, although well known since the early developments following Berry’s discovery, has not been widely applied in the context considered here. The connection between our approach and the asymptotic method proposed by Strominger and collaborators appears \cite{stro,stro1,stro2,stro3}, in principle, natural, since both rely on effective theories. However, the exact equivalence between the two descriptions is not yet established; it is plausible that they represent reciprocal versions of one another—perhaps related through a kind of Fourier transform—but the precise relation remains an open issue under discussion.}

\acknowledgments
\noindent 
One of us (JG) is pleased to acknowledge Prof. Pierre Sikivie for valuable discussions and comments at the early stages of this investigation. This research was supported by DICYT (USACH), grant number 042531GR\_REG and 042531MF\_REG.
\appendix
\section{Derivation of the Modified Planck Distribution}

We show how a functional Berry phase accumulated by electromagnetic radiation modes can induce an oscillatory correction to Planck's law for the blackbody spectrum. Starting from the expectation value of the energy operator, we analyze the coherent contribution between states \( |n\rangle \) and \( |m\rangle \), modulated by a functional phase \( \Delta \alpha_n \). Within the leading-order infrared (soft-photon) approximation, we obtain a corrected spectral energy density of the form:

\begin{equation}
\rho(\nu) = \rho_{\text{Planck}}(\nu) \left[1 + \epsilon \cos\left( \Delta \alpha(\nu) \right)\right],
\end{equation}
where \( \epsilon \ll 1 \) and \( \Delta \alpha(\nu) \) is a frequency-dependent phase.

We consider a quantized electromagnetic field inside a cavity. The energy operator for a mode of frequency \( \nu \) is given by:
\begin{equation}
\hat{E}_\nu = h\nu\, a^\dagger_\nu a_\nu.
\end{equation}

Suppose the system is in a state expressed as a superposition in the number basis:
\begin{equation}
|\Psi\rangle = \sum_n c_n\, e^{i \Delta \alpha_n} |n\rangle,
\end{equation}
where \( \Delta \alpha_n \) is a functional phase that depends on the mode and is related to infrared coupling and the functional Berry connection. Importantly, in the adiabatic approximation we can use:
\begin{equation}
\Delta \alpha_n \approx n\, \Delta \alpha.
\end{equation}

The expectation value of the energy is then:
\begin{align}
\langle \hat{E}_\nu \rangle 
&= \sum_{n,m} c_n^* c_m\, e^{-i\Delta \alpha_n} e^{i\Delta \alpha_m} \bra{n} h\nu\, a^\dagger_\nu a_\nu \ket{m} \notag \\
&= h\nu \sum_n |c_n|^2\, n + h\nu \sum_{n\neq m} c_n^* c_m \bra{n} a^\dagger_\nu a_\nu \ket{m}\, e^{i(\Delta \alpha_m - \Delta \alpha_n)}.
\end{align}

The second sum contains coherent terms, which typically vanish in thermal distributions. However, if partial coherence exists between \( |n\rangle \) and \( |n+1\rangle \), then:
\begin{equation}
\bra{n} a^\dagger a \ket{n+1} = \sqrt{(n+1)}\, \delta_{n+1,m}.
\end{equation}

This leads to a contribution of the form:
\begin{align}
\langle \hat{E}_\nu \rangle 
&\approx h\nu \sum_n |c_n|^2\, n 
+ h\nu \sum_n \Re\left( c_n^* c_{n+1}\, \sqrt{n+1}\, e^{i(\Delta \alpha_{n+1} - \Delta \alpha_n)} \right).
\end{align}

Assuming the functional phase varies smoothly with \( n \), we can write:
\begin{equation}
\Delta \alpha_{n+1} - \Delta \alpha_n \approx {\Delta \alpha}\,{n} \approx \Delta \alpha(\nu),
\end{equation}
and suppose \( c_n^* c_{n+1} \sqrt{n+1} \) is real and positive (quasi-coherent state). Then:
\begin{equation}
\langle \hat{E}_\nu \rangle \approx \langle \hat{E}_\nu \rangle_{\text{Planck}} \left[1 + \epsilon \cos\left( \Delta \alpha(\nu) \right)\right],
\end{equation}
where \( \epsilon \ll 1 \) quantifies the relative amplitude of the oscillations.

The spectral energy density is defined as:
\begin{equation}
\rho(\nu) = \frac{\langle \hat{E}_\nu \rangle}{V\, d\nu},
\end{equation}
and for an ideal blackbody it is given by Planck’s law:
\begin{equation}
\rho_{\text{Planck}}(\nu) = \frac{8\pi h \nu^3}{c^3} \frac{1}{e^{h\nu/kT} - 1}.
\end{equation}

Therefore, the functional correction can be written as:
\begin{equation}
\rho(\nu) = \rho_{\text{Planck}}(\nu) \left[1 + \epsilon \cos\left( \Delta \alpha(\nu) \right)\right],
\end{equation}
where \( \Delta \alpha(\nu) \) typically takes the form:
\begin{equation}
\Delta \alpha(\nu) = \frac{2\pi \nu}{\Lambda_{\text{IR}}} + \phi,
\end{equation}
with \( \Lambda_{\text{IR}} \) an infrared scale associated with the functional coupling, and \( \phi \) a constant phase.

The presence of functional phases associated with the Berry connection in radiation modes can induce oscillations on top of Planck’s law. These oscillations constitute a possible observable signature of nonperturbative infrared effects in gauge theories.

%

%
%
%
%
%
%
%
%
%
%
%
%
%


\begin{thebibliography}{99}
\bibitem{BN}
F. Bloch and A. Nordsieck, ``Note on the Radiation Field of the Electron'', \emph{Phys. Rev.} \textbf{52}, 54 (1937).

\bibitem{YFS}
D. R. Yennie, S. C. Frautschi, and H. Suura, ``The infrared divergence phenomena and high-energy processes'', \emph{Annals Phys.} \textbf{13}, 379 (1961).

\bibitem{KI}
T. Kinoshita, ``Mass Singularities of Feynman Amplitudes'', \emph{J. Math. Phys.} \textbf{3}, 650 (1962).

\bibitem{LN}
T. D. Lee and M. Nauenberg, ``Degenerate Systems and Mass Singularities'', \emph{Phys. Rev.} \textbf{133}, B1549 (1964).

\bibitem{weinberg} S.~Weinberg,
``Infrared photons and gravitons,''
Phys. Rev. \textbf{140} (1965), B516-B524
doi:10.1103/PhysRev.140.B516.
\bibitem{chung} See also V.~Chung,
``Infrared Divergence in Quantum Electrodynamics,''
Phys. Rev. \textbf{140} (1965), B1110-B1122
doi:10.1103/PhysRev.140.B1110.

\bibitem{KF}
P. P. Kulish and L. D. Faddeev, ``Asymptotic conditions and infrared divergences in quantum electrodynamics'', \emph{Theor. Math. Phys.} \textbf{4}, 745 (1970).

\bibitem{carney1} D.~Carney, L.~Chaurette, D.~Neuenfeld and G.~W.~Semenoff,
``Infrared quantum information,''
Phys. Rev. Lett. \textbf{119} (2017) no.18, 180502
doi:10.1103/PhysRevLett.119.180502
[arXiv:1706.03782 [hep-th]].

\bibitem{carney2} D.~Carney, L.~Chaurette, D.~Neuenfeld and G.~W.~Semenoff , 
``Dressed infrared quantum information,''
Phys. Rev. D \textbf{97} (2018) no.2, 025007
doi:10.1103/PhysRevD.97.025007
[arXiv:1710.02531 [hep-th]].




\bibitem{ref1} D.~A.~Forde and A.~Signer,
``Infrared finite amplitudes for massless gauge theories,''
Nucl. Phys. B \textbf{684} (2004), 125-161
doi:10.1016/j.nuclphysb.2004.02.024
[arXiv:hep-ph/0311059 [hep-ph]].

\bibitem{ref2} A.~Misra,
``Coherent states in null plane QED.,''
Phys. Rev. D \textbf{50} (1994), 4088-4096
doi:10.1103/PhysRevD.50.4088
[arXiv:hep-th/9311101 [hep-th]].

\bibitem{reclamo1} C.~Gomez, R.~Letschka and S.~Zell,
``Infrared Divergences and Quantum Coherence,''
Eur. Phys. J. C \textbf{78} (2018) no.8, 610
doi:10.1140/epjc/s10052-018-6088-2
[arXiv:1712.02355 [hep-th]]; C.~G{\'o}mez, R.~Letschka and S.~Zell,
``The Scales of the Infrared,''
JHEP \textbf{09} (2018), 115
doi:10.1007/JHEP09(2018)115
[arXiv:1807.07079 [hep-th]].

\bibitem{reclamo2} J.~D.~More and A.~Misra,
``Infra-red Divergences in Light-Front QED and Coherent State Basis,''
Phys. Rev. D \textbf{86} (2012), 065037
doi:10.1103/PhysRevD.86.065037
[arXiv:1206.3097 [hep-th]]; J.~D.~More and A.~Misra,
``Cancellation of infrared divergences to all orders in light front QED,''
Phys. Rev. D \textbf{89} (2014) no.10, 105021
doi:10.1103/PhysRevD.89.105021
[arXiv:1402.4924 [hep-th]].
\bibitem{FS}
J. Fröhlich, G. Morchio, and F. Strocchi, ``Infrared problem and spontaneous breaking of the Lorentz group in QED'', \emph{Phys. Lett. B} \textbf{89}, 61 (1979).

\bibitem{MorchioStrocchi}
G.~Morchio and F.~Strocchi,
``Infrared singularities, vacuum structure and pure phases in local quantum field theory,''
\emph{Ann. Inst. H. Poincaré Phys. Théor.} \textbf{33}, 251–282 (1980).

\bibitem{StrocchiBook}
F.~Strocchi, 
\emph{Symmetry Breaking}, 2nd ed., Lecture Notes in Physics, vol. 732, Springer (2008). Chapter 5.

\bibitem{DerezinskiGerard}
J.~Dereziński and C.~Gérard, 
\emph{Mathematics of Quantization and Quantum Fields}, Cambridge University Press (2013). Chapters 10–13.

\bibitem{Buchholz1}
D.~Buchholz, 
``The physical state space of quantum electrodynamics,'' 
\emph{Commun. Math. Phys.} \textbf{85}, 49–71 (1982).

\bibitem{Buchholz2}
D.~Buchholz, 
``Gauss’ law and the infraparticle problem,'' 
\emph{Phys. Lett. B} \textbf{174}, 331–334 (1986).

\bibitem{Mund}
J.~Mund,
``No-Go Theorem for Free Relativistic Anyons in $d = 2 + 1$,'' 
\emph{Lett. Math. Phys.} \textbf{43}, 319–328 (1998).

\bibitem{Herbst}
I.~Herbst and E.~Skibsted, 
``Quantum scattering for N-body long-range systems at thresholds,'' 
\emph{Adv. Math.} \textbf{270}, 138–218 (2015). \href{https://doi.org/10.1016/j.aim.2014.09.015}{doi:10.1016/j.aim.2014.09.015}


\bibitem{Herdegen}
A. Herdegen, ``Asymptotic structure of electrodynamics revisited'', \emph{Lett. Math. Phys.} \textbf{81}, 99–113 (2007).


\bibitem{Berry}
M.~V.~Berry,
``Quantal phase factors accompanying adiabatic changes,''
Proc. Roy. Soc. Lond. A \textbf{392} (1984), 45-57

\bibitem{yo}
J.~Gamboa,
``Berry Phase in Non-Perturbative QED,''
[arXiv:2503.24194 [hep-th]]; ``Entanglement and Effective Field Theories,''
[arXiv:2502.11819 [hep-th]]. To appear in PLB.

\bibitem{yo2} J.~Gamboa,
``Topology and the Infrared Structure of Quantum Electrodynamics,''
[arXiv:2505.13247 [hep-th]]. To appear in JHEP.

\bibitem{shapere} A. ~Shapere and F~.~Wilczek, ``Geometric Phases in Physics'', World Scientific (1989), pag. 160.

\bibitem{stro}
A. Strominger, ``Lectures on the Infrared Structure of Gravity and Gauge Theory'', \href{https://arxiv.org/abs/1703.05448}
{arXiv:1703.05448}.


\bibitem{stro1}
A.~Strominger, ``On BMS Invariance of Gravitational Scattering,''  
\textit{JHEP} \textbf{07} (2014) 152 [arXiv:1312.2229].

\bibitem{stro2}
A.~Strominger and A.~Zhiboedov, ``Gravitational Memory, BMS Supertranslations and Soft Theorems,''  
\textit{JHEP} \textbf{01} (2016) 086 [arXiv:1411.5745].

\bibitem{stro3}
T.~He, P.~Mitra, A.~P.~Porfyriadis, and A.~Strominger, ``New Symmetries of Massless QED,''  
\textit{JHEP} \textbf{10} (2014) 112 [arXiv:1407.3789].


\bibitem{fujikawa1} K.~Fujikawa,
``Path Integral for Gauge Theories with Fermions,''
Phys. Rev. D \textbf{21} (1980), 2848
[erratum: Phys. Rev. D \textbf{22} (1980), 1499]
doi:10.1103/PhysRevD.21.2848.

\bibitem{fujikawa2} K. Fujikawa, ``Path Integral Measure for Gauge Invariant Fermion Theories,''
Phys. Rev. Lett. \textbf{42} (1979), 1195-1198
doi:10.1103/PhysRevLett.42.1195.

\bibitem{Atiyah1}
M.~F.~Atiyah and I.~M.~Singer,
``Dirac Operators Coupled to Vector Potentials,''
\textit{Proc. Natl. Acad. Sci. USA} \textbf{81}, 2597 (1984).

\bibitem{Witten1}
E.~Witten,
``Fermion Path Integrals and Topological Phases,''
\textit{Rev. Mod. Phys.} \textbf{88}, 035001 (2016),
\href{https://arxiv.org/abs/1508.04715}{arXiv:1508.04715 [hep-th]}.

\bibitem{Hasan}
M.~Z.~Hasan and C.~L.~Kane, 
``\href{https://doi.org/10.1103/RevModPhys.82.3045}{Colloquium: Topological insulators},''
\emph{Rev. Mod. Phys.} \textbf{82}, 3045 (2010).

\bibitem{Qi}
X.-L.~Qi and S.-C.~Zhang, 
``\href{https://doi.org/10.1103/RevModPhys.83.1057}{Topological insulators and superconductors},''
\emph{Rev. Mod. Phys.} \textbf{83}, 1057 (2011).

\bibitem{Moore}
J.~E.~Moore, 
``\href{https://doi.org/10.1038/nature08916}{The birth of topological insulators},''
\emph{Nature} \textbf{464}, 194–198 (2010).

\bibitem{Kane}
C.~L.~Kane and E.~J.~Mele,
``\href{https://doi.org/10.1103/PhysRevLett.95.146802}{Quantum Spin Hall Effect in Graphene},''
\emph{Phys. Rev. Lett.} \textbf{95}, 146802 (2005).

\bibitem{Berne}
B.~A.~Bernevig, T.~L.~Hughes, and S.-C.~Zhang,
``\href{https://doi.org/10.1126/science.1133734}{Quantum Spin Hall Effect and Topological Phase Transition in HgTe Quantum Wells},''
\emph{Science} \textbf{314}, 1757 (2006).




\bibitem{ir1}
D.~J. Fixsen, ``The Temperature of the Cosmic Microwave Background,'' 
\emph{Astrophysical Journal}, vol.~707, no.~2, pp.~916--920, 2009. 
\href{https://doi.org/10.1088/0004-637X/707/2/916}{doi:10.1088/0004-637X/707/2/916}.

\bibitem{ir2}
J.~C. Mather, E.~S. Cheng, D.~A. Cottingham, R.~E. Eplee Jr., D.~J. Fixsen, T.~Hewagama, R.~B. Isaacman, S.~S. Meyer, P.~D. Noerdlinger, R.~A. Shafer, E.~L. Wright, C.~L. Bennett, N.~W. Boggess, E.~D. Jackson, S.~M. Silverberg, K.~A. Turpie, and S.~T. Wilkinson,  
``Measurement of the Cosmic Microwave Background Spectrum by the COBE FIRAS Instrument,''  
\emph{Astrophysical Journal}, vol.~420, pp.~439--444, 1994.  
\href{https://doi.org/10.1086/173574}{doi:10.1086/173574}.

\bibitem{ir3}
Planck Collaboration (P.~A.~R. Ade et al.),  
``Planck 2013 results. XVI. Cosmological parameters,''  
\emph{Astronomy \& Astrophysics}, vol.~571, A16, 2014.  
\href{https://doi.org/10.1051/0004-6361/201321591}{doi:10.1051/0004-6361/201321591}.




\bibitem{Fixsen:1996}
D.~J.~Fixsen et al.,
``The Cosmic Microwave Background spectrum from the full COBE FIRAS data set,''
Astrophys. J. \textbf{473}, 576 (1996).
doi:10.1086/178173

\bibitem{WMAP:2013}
G.~Hinshaw et al. [WMAP Collaboration],
``Nine-Year Wilkinson Microwave Anisotropy Probe (WMAP) Observations: Cosmological Parameter Results,''
Astrophys. J. Suppl. \textbf{208}, 19 (2013).
doi:10.1088/0067-0049/208/2/19

\bibitem{Planck:2018}
N.~Aghanim et al. [Planck Collaboration],
``Planck 2018 results. VI. Cosmological parameters,''
Astron. Astrophys. \textbf{641}, A6 (2020).
doi:10.1051/0004-6361/201833910

\bibitem{QUIJOTE:2020}
R.~Genova-Santos et al. [QUIJOTE Collaboration],
``The QUIJOTE experiment: project overview and first results,''
Mon. Not. R. Astron. Soc. \textbf{495}, 975 (2020).
doi:10.1093/mnras/staa1208

\bibitem{QUIJOTE:2023}
J.~M.~Herranz et al. [QUIJOTE Collaboration],
``QUIJOTE scientific results and future perspectives,''
Galaxies \textbf{11}, 33 (2023).
doi:10.3390/galaxies11020033

\bibitem{GamboaRJ}
J.~Gamboa,
``Topological Relics of Infrared QED in the Rayleigh--Jeans Regime,''
arXiv:2505.13247 [hep-th] (2025).

\bibitem{Simon}
B.~Simon,
``Holonomy, the quantum adiabatic theorem, and Berry's phase,''
Phys. Rev. Lett. \textbf{51} (1983), 2167-2170
doi:10.1103/PhysRevLett.51.2167

\end{thebibliography}
\end{document}